\documentclass[11pt, leqno]{article}
\usepackage{times, amsfonts, amsmath, latexsym, graphicx}

\newtheorem{theorem}{Theorem}[section]
\newtheorem{lemma}[theorem]{Lemma}
\newtheorem{corollary}[theorem]{Corollary}
\newtheorem{example}[theorem]{Example}

\newenvironment{Lemma}
	{\begin{lemma}\sl}
	{\end{lemma}}

\newenvironment{Example}
	{\begin{example}\rm}
	{\end{example}}

\def\Pr{\mathrm{P}}
\def\Ex{\mathrm{E}\,}

\def\bs{\boldsymbol}

\def\a{\bs{a}}
\def\A{\bs{A}}
\def\b{\bs{b}}
\def\B{\bs{B}}
\def\e{\bs{e}}
\def\H{\bs{H}}
\def\x{\bs{x}}

\def\eps{\varepsilon}
\def\sig{\sigma}
\def\th{\theta}
\def\thb{\bs{\theta}}
\def\etab{\bs{\eta}}

\def\R{\mathbb{R}}

\def\K{\mathbb{K}}
\def\V{\mathbb{V}}

\def\CC{\mathcal{C}}

\def\EE{\mathcal{E}}
\def\UU{\mathcal{U}}
\def\VV{\mathcal{V}}

\def\argmin{\mathop{\rm argmin}}

\def\bea{\begin{eqnarray*}}
\def\eea{\end{eqnarray*}}

\newcommand{\ruck}[1]{\strut\hspace{#1cm}}

\begin{document}

\addtolength{\baselineskip}{+.5\baselineskip}
\addtolength{\parskip}{+.15\baselineskip}

\thispagestyle{empty}

\title{Least Squares and Shrinkage Estimation under Bimonotonicity
Constraints}
\author{Rudolf Beran and Lutz D\"umbgen\\
	University of California at Davis and University of Bern}
\date{September 2008, revised January 2009}

\maketitle

\begin{abstract}
In this paper we describe active set type algorithms for minimization of a smooth function under general order constraints, an important case being functions on the set of bimonotone $r \times s$ matrices. These algorithms can be used, for instance, to estimate a bimonotone regression function via least squares or (a smooth approximation of) least absolute deviations. Another application is shrinkage estimation in image denoising or, more generally, regression problems with two ordinal factors after representing the data in a suitable basis which is indexed by pairs $(i,j) \in \{1,\ldots,r\} \times \{1,\ldots,s\}$. Various numerical examples illustrate our methods.
\end{abstract}



\section{Introduction}

Monotonicity and other qualitative constraints play an important role in contemporary nonparametric statistics. One reason for this success is that such constraints are often plausible or even justified theoretically, within an appropriate mathematical formulation of the application. Moreover, by imposing shape constraints one can often avoid more traditional smoothness assumptions which typically lead to procedures requiring the choice of some tuning parameter. A good starting point for statistical inference under qualitative constraints is the monograph by Robertson et al.\ \cite{robertson_wright_dykstra_1988}.

Estimation under order constraints leads often to the following optimization problem: For some dimension $p \ge 2$ let $Q : \R^p \to \R$ be a given functional. For instance,
\begin{equation}
\label{eq: WLS}
	Q(\thb) \ = \ \sum_{u=1}^p w_u (Z_u - \th_u)^2
\end{equation}
with a certain weight vector $\bs{w} \in (0,\infty)^p$ and a given data
vector $\bs{Z} \in \R^p$. In general we assume that $Q$ is continuously
differentiable, strictly convex and coercive, i.e.
$$
	Q(\thb) \ \to \ \infty
	\quad\text{as} \ \|\thb\| \to \infty ,
$$ 
where $\|\cdot\|$ is some norm on $\R^p$. The goal is to minimize $Q$ over the following subset $\K$ of $\R^p$: Let $\CC$ be a given collection of pairs $(u,v)$ of different indices $u,v \in \{1,2,\ldots,p\}$, and define
$$
	\K = \K(\CC)
	\ = \ \bigl\{ \thb \in \R^p :
		\th_u \le \th_v \ \text{for all} \ (u,v) \in \CC \bigr\} .
$$
This defines a closed convex cone in $\R^p$ containing all constant
vectors.

For instance, if $\CC$ consists of $(1,2)$, $(2,3)$, \ldots, $(p-1,p)$, then $\K$ is the cone of all vectors $\thb \in \R^p$ such that $\th_1 \le \th_2 \le \cdots \le \th_p$. Minimizing (\ref{eq: WLS}) over all such vectors is a standard problem and can be solved in $O(p)$ steps via the pool-adjacent-violators algorithm (PAVA). The latter was introduced in a special setting by Ayer et al.\ \cite{ayer_brunk_wright_silverman_1955} and extended later by numerous authors, see \cite{robertson_wright_dykstra_1988} and Best and Chakravarti \cite{best_chakravarti_1990}.

As soon as $Q(\cdot)$ is not of type (\ref{eq: WLS}) or $\CC$ differs from the aforementioned standard example, the minimization of $Q(\cdot)$ over $\K$ becomes more involved. Here is another example for $\K$ and $\CC$ which is of primary interest in the present paper: Let $p = rs$ with integers $r,s \ge 2$, and identify $\R^p$ with the set $\R^{r\times s}$ of all matrices with $r$ rows and $s$ columns. Further let $\K_{r,s}$ be the set of all matrices $\thb \in \R^{r\times s}$ such that
$$
	\th_{i,j} \le \th_{i+1,j} \ \text{whenever} \ i < r
	\quad\text{and}\quad
	\th_{i,j} \le \th_{i,j+1} \ \text{whenever} \ j < s .
$$
This corresponds to the set $\CC_{r,s}$ of all pairs $\bigl( (i,j),
(k,\ell) \bigr)$ with $i,k \in \{1,\ldots,r\}$ and $j,\ell \in
\{1,\ldots,s\}$ such that either $(k,\ell) = (i+1,j)$ or $(k,\ell) =
(i,j+1)$. Hence there are $\# \CC = 2rs - r - s$ constraints.

Minimizing the special functional (\ref{eq: WLS}), i.e.\ $Q(\thb) =
\sum_{i,j} w_{ij}(Z_{ij} - \th_{ij})^2$, over the bimonotone cone
$\K_{r,s}$ is a well recognized problem with various proposed solutions,
see, for instance, Spouge et al.\ \cite{spouge_wan_wilbur_2003}, Burdakow
et al.\ \cite{burdakow_grimwall_hussian_2004}, and the references cited
therein. However, all these algorithms exploit the special structure of
$\K_{r,s}$ or (\ref{eq: WLS}). For general functionals $Q(\cdot)$, e.g.\
quadratic functions with positive definite but non-diagonal hessian matrix,
different approaches are needed.

The remainder of this paper is organized as follows. In Section~\ref{sec: WLS for bimonotone regression} we describe the \textsl{bimonotone regression} problem and argue that the special structure (\ref{eq: WLS}) is sometimes too restrictive even in that context. In Section~\ref{sec: algorithm} we derive possible algorithms for the general optimization problem described above. These algorithms involve a discrete optimization step which gives rise to a dynamic program in case of $\K = \K_{r,s}$. For a general introduction to dynamic programming see Cormen et al.\ \cite{cormen_leiserson_rivest_1990}.
Other ingredients are active methods as described by, for instance, Fletcher \cite{fletcher_1987}, Best and Chakravarti \cite{best_chakravarti_1990} or D\"umbgen et al.\ \cite{duembgen_huesler_rufibach_2007}, sometimes combined with the ordinary PAVA in a particular fashion. It will be shown that all these algorithms find the exact solution in finitely many steps, at least when $Q(\cdot)$ is an arbitrary quadratic and strictly convex function. Finally, in Section~\ref{sec: shrinkage} we adapt our procedure to image denoising via \textsl{bimonotone shrinkage} of generalized Fourier coefficients. The statistical method in this section was already indicated in Beran and D\"umbgen \cite{beran_duembgen_1998} but has not been implemented yet, for lack of an efficient computational algorithm.

\section{Least squares estimation of bimonotone regression functions}
\label{sec: WLS for bimonotone regression}

Suppose that one observes $(x^1,y^1,Z^1)$, $(x^2,y^2,Z^2)$, \ldots,
$(x^n,y^n,Z^n)$ with real components $x^t$, $y^t$ and $Z^t$. The points
$(x^t,y^t)$ are regarded as fixed points, which is always possible by
conditioning, while
$$
	Z^t \ = \ \mu(x^t,y^t) + \eps^t
$$
for an unknown regression function $\mu : \R\times\R \to \R$ and independent random errors $\eps^1$, $\eps^2$, \ldots, $\eps^n$ with mean zero. In some applications it is plausible to assume $\mu$ to be bimonotone increasing, i.e.\ non-decreasing in both arguments. Then it would be desirable to estimate $\mu$ under that constraint only. One possibility would be to
minimize
$$
	\sum_{t=1}^n (Z^t - \mu(x^t,y^t))^2
$$
over all bimonotone functions $\mu$. The resulting minimizer $\hat{\mu}$ is
uniquely defined on the finite set of all design points $(x^t,y^t)$, $1 \le
t \le n$.

For a more detailed discussion, suppose that we want to estimate $\mu$ on a finite rectangular grid
$$
	\bigl\{ (x_{(i)}, y_{(j)}) : 1 \le i \le r, 1 \le j \le s \bigr\} ,
$$
where $x_{(1)} < x_{(2)} < \cdots < x_{(r)}$ and $y_{(1)} < y_{(2)} < \cdots < y_{(s)}$ contain at least the different elements of $\{x^1, x^2, \ldots, x^n\}$ and $\{y^1,y^2,\ldots,y^n\}$, respectively, but maybe additional points as well. For $1 \le i \le r$ and $1 \le j \le s$ let $w_{ij}$ be the number of all $t \in \{1,\ldots,n\}$ such that $(x^t,y^t) = (x_{(i)},y_{(j)})$, and let $Z_{ij}$ be the average of $Z^t$ over these indices $t$. Then $\sum_{t=1}^n (Z^t - \mu(x^t,y^t))^2$ equals
$$
	Q(\thb) \ = \ \sum_{i,j} w_{ij}(Z_{ij} - \theta_{ij})^2 ,
$$
where $\thb = (\theta_{ij})_{i,j}$ stands for the matrix $\bigl( \mu(x_{(i)},y_{(j)}) \bigr)_{i,j} \in \K_{r,s}$.

\paragraph{Setting 1: Complete layout.}
Suppose that $w_{ij} > 0$ for all $(i,j) \in \{1,\ldots,r\} \times \{1,\ldots,s\}$. Then the resulting optimization problem is precisely the one described in the introduction.

\paragraph{Setting 2a: Incomplete layout and simple interpolation/extrapolation.} 
Suppose that the set $\UU$ of all index pairs $(i,j)$ with $w_{ij} > 0$ differs from $\{1,\ldots,r\} \times \{1,\ldots,s\}$. Then
$$
	Q(\thb) \ = \ \sum_{u \in \UU} w_u(Z_u - \theta_u)^2
$$
fails to be coercive. Nevertheless it can be minimized over $\K_{r,s}$ with the algorithms described later. Let $\check{\thb}$ be such a minimizer. Since it is uniquely defined on $\UU$ only, we propose to replace it with $\hat{\thb} = 2^{-1}(\underline{\thb} + \overline{\thb})$, where
\bea
	\underline{\theta}_{ij}
	& = & \max \Bigl( \bigl\{ \check{\theta}_{i'j'} :
			(i',j') \in \UU, i' \le i, j' \le j \bigr\}
		\cup \{\check{\theta}_{\rm min}\} \Bigr) , \\
	\overline{\theta}_{ij}
	& = & \min \Bigl( \bigl\{ \check{\theta}_{i'j'} :
			(i',j') \in \UU, i \le i', j \le j' \bigr\}
		\cup \{\check{\theta}_{\rm max}\} \Bigr) ,
\eea
and $\check{\theta}_{\rm min}$ and $\check{\theta}_{\rm max}$ denote the minimum and maximum, respectively, of $\{\check{\theta}_u : u \in \UU\}$. Note that $\underline{\thb}$ and $\overline{\thb}$ belong to $\K_{r,s}$ and are extremal in the sense that any matrix $\thb \in \K_{r,s} \cap [\check{\theta}_{\rm min}, \check{\theta}_{\rm max}]^{r \times s}$ with $\theta_u = \check{\theta}_u$ for all $u \in \UU$ satisfies necessarily $\underline{\theta}_{ij} \le \theta_{ij}^{} \le \overline{\theta}_{ij}$ for all $(i,j)$.

\paragraph{Setting 2b: Incomplete layout and light regularization.}
Instead of restricting one's attention to the index set $\UU$, one can estimate the full matrix $\bigl( \mu(x_{(i)},y_{(j)}) \bigr)_{i,j} \in \R^{r\times s}$ by minimizing a suitably penalized sum of squares,
$$
	Q(\thb) \ = \ \sum_{u \in \mathcal{U}} w_u(Z_u - \th_u)^2
		+ \lambda P(\thb) ,
$$
over $\K_{r,s}$ for some small parameter $\lambda > 0$. Here $P(\cdot)$ is a convex quadratic function on $\R^{r\times s}$ such that $Q(\cdot)$ is strictly convex. One possibility would be Tychonov regularisation with $P(\thb) = \sum_{i,j} (\theta_{ij} - \theta_o)^2$ and a certain reference value $\th_o$, for instance, $\theta_o = \sum_{i,j} w_{ij} Z_{ij} \big/ \sum_{i,j} w_{ij}$. In our particular setting we prefer the penalty
\begin{equation}
\label{eq: interpolation penalty}
	P(\thb) \ = \ \sum_{((i,j), (k,\ell)) \, \in \, \CC_{r,s}}
			(\th_{k\ell} - \th_{ij})^2 ,
\end{equation}
because it yields smoother interpolations than the recipe for Setting~2a or the Tychonov penalty. One can easily show that the resulting quadratic function $Q$ is strictly convex but with non-diagonal hessian matrix. Thus it fulfills our general requirements but is not of type (\ref{eq: WLS}).

Note that adding a penalty term such as (\ref{eq: interpolation penalty}) could be worthwhile even in case of a complete layout if the underlying function $\mu$ is assumed to be smooth. But this leads to the nontrivial task of choosing $\lambda > 0$ appropriately. Here we use the penalty term mainly for smooth interpolation/extrapolation with $\lambda$ just large enough to ensure a well-conditioned Hessian matrix. We refer to this as ``light regularization'', and the exact value of $\lambda$ is essentially irrelevant.

\begin{figure}[h]
\includegraphics[width=0.49\textwidth]{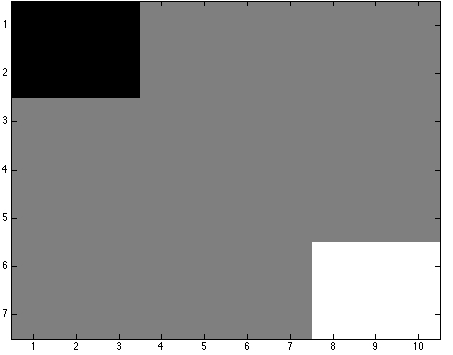}
\includegraphics[width=0.49\textwidth]{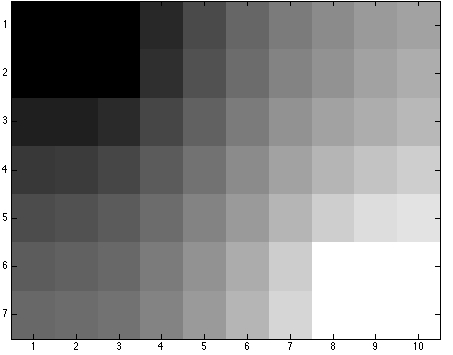}
\caption{Simple interpolation/extrapolation versus light regularization}
\label{fig: InterpolateEx}
\end{figure}

\begin{Example}
To illustrate the difference between simple interpolation/extrapolation and light regularization with penalty (\ref{eq: interpolation penalty}) we consider just two observations, $(x^1,y^1,Z^1) = (2,3,0)$ and $(x^2,y^2,Z^2) = (6,7,1)$, and let $r = 7$, $s = 10$ with $x_{(i)} = i$ and $y_{(j)} = j$. Thus $w_{ij} = 0$ except for $w_{2,3} = w_{6,7} = 1$, while $Z_{2,3} = 0$ and $Z_{6,7} = 1$. Any minimizer $\check{\thb}$ of $\sum_{u \in \UU} w_u(Z_u - \th_u)^2$ over $\K_{7,10}$ satisfies $\check{\theta}_{2,3} = 0$ and $\check{\theta}_{6,7} = 1$, so the recipe for Setting~2a yields
$$
	\hat{\theta}_{ij} \ = \ \begin{cases}
		0 & \text{if} \ i \le 2, j \le 3 , \\
		1 & \text{if} \ i \ge 6, j \ge 7 , \\
		0.5 & \text{else} .
	\end{cases} 
$$
The left panel of Figure~\ref{fig: InterpolateEx} shows the latter fit $\hat{\thb}$, while the right panel shows the regularized fit based on (\ref{eq: interpolation penalty}) with $\lambda = 10^{-4}$. In these and most subsequent pictures we use a gray scale from $\textrm{black} = 0$ to $\textrm{white} = 1$.
\end{Example}

\begin{Example} \textbf{(Binary regression).} \
\label{ex: binary regression}
We generated a random matrix $\bs{Z} \in \{0,1\}^{r\times s}$ with $r = 70$
rows, $s = 100$ columns and independent components $Z_{ij}$, where
$$
	\Pr(Z_{ij} = 1) \ = \ \theta_{ij}
		= \frac{x_{(i)} + y_{(j)}}{4}
			+ \frac{1 \bigl\{ y_{(j)} \ge 1/2 + \cos(\pi x_{(i)})/4 \bigr\}}{2}
$$
with $x_{(i)} = (i - 0.5)/r$ and $y_{(j)} = (j - 0.5)/s$. Thereafter we removed randomly all but $700$ of the $7000$ components $Z_{ij}$. The resulting data are depicted in the upper left panel of Figure~\ref{fig: binary regression incomplete}, where missing values are depicted grey, while the upper right panel shows the true signal $\thb$. The lower panels depict the least squares estimator with simple interpolation/extrapolation (left) and light regularization based on (\ref{eq: interpolation penalty}) with $\lambda = 10^{-4}$ (right). Note that both estimators are very similar. Due to the small value of $\lambda$, the main differences occur in regions without data points.

\begin{figure}[h]
\includegraphics[width=0.49\textwidth]{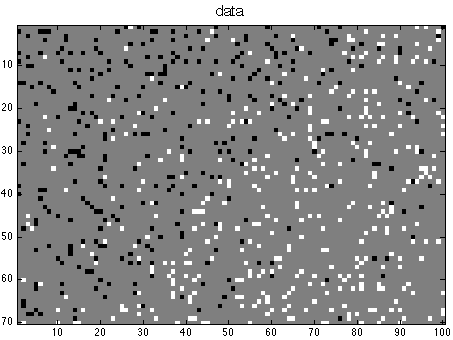}
\includegraphics[width=0.49\textwidth]{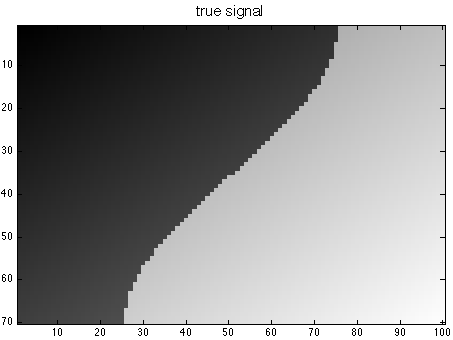}

\includegraphics[width=0.49\textwidth]{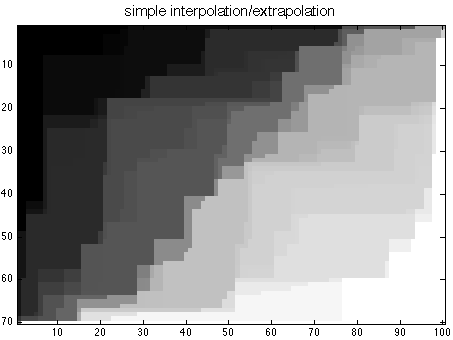}
\includegraphics[width=0.49\textwidth]{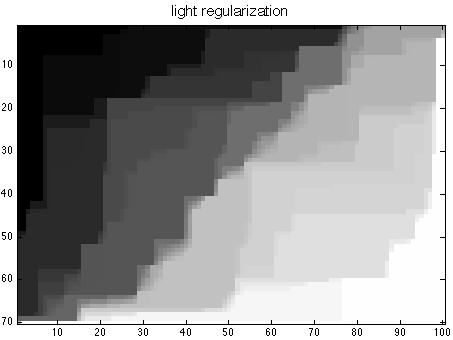}
\caption{Binary regression with incomplete layout}
\label{fig: binary regression incomplete}
\end{figure}

The quality of an estimator $\hat{\thb}$ for $\thb$ may be quantified by the average absolute deviation,
$$
	\mathrm{AAD}
	\ = \ \frac{1}{rs} \sum_{i=1}^r \sum_{j=1}^s |\hat{\th}_{ij} - \th_{ij}| .
$$
For the estimator with simple interpolation/extrapolation, $\mathrm{AAD}$ turned out to be $7.5607 \cdot 10^{-2}$, the estimator based on light regularization performed slightly better with $\mathrm{AAD} = 7.4039 \cdot 10^{-2}$.
\end{Example}

\section{The general algorithmic problem}
\label{sec: algorithm}

We return to the general framework introduced in the beginning with a
continuously differentiable, strictly convex and coercive functional $Q :
\R^p \to \R$ and a closed convex cone $\K = \K(\CC) \in \R^p$ determined by
a collection $\CC$ of inequality constraints.

Before starting with explicit algorithms, let us characterize the point
$$
	\hat{\thb} \ = \ \argmin_{\thb \in \K} \, Q(\thb) .
$$
It is well-known from convex analysis that a point $\thb \in \K$ coincides
with $\thb$ if, and only if,
\begin{equation}
\label{ineq: characterize thetahat}
	\nabla Q(\thb)^\top \thb \ = \ 0 \ \le \ \nabla Q(\thb)^\top \etab
	\quad\text{for all} \ \etab \in \K ,
\end{equation}
where $\nabla Q(\thb)$ denotes the gradient of $Q$ at $\thb$. This characterization involves infinitely many inequalities, but it can be replaced with a criterion involving only finitely many constraints.

\subsection{Extremal directions of $\K$}
\label{subsec: extremal set}

Note that $\K$ contains all constant vectors $c \bs{1}$, $c \in \R$, where $\bs{1} = \bs{1}_p = (1)_{i=1}^p$. It can be represented as follows:

\begin{Lemma}
\label{lem: extremal set}
Define
$$
	\EE \ = \ \K \cap \{0,1\}^p .
$$
Then any vector $\x \in \K$ may be represented as
$$
	\x \ = \ \min(\x) \bs{1}
		+ \sum_{\bs{e} \in \EE} \lambda_{\e} \e
$$
with coefficients $\lambda_{\e} \ge 0$ such that $\sum_{\e \in \EE}
\lambda_{\e} = \max(\x) - \min(\x)$.
\end{Lemma}

Here $\min(\x)$ and $\max(\x)$ denote the minimum and maximum,
respectively, of the components of $\x$.

\paragraph{Modified characterization of $\hat{\thb}$.}
By means of Lemma~\ref{lem: extremal set} one can easily verify that
(\ref{ineq: characterize thetahat}) is equivalent to the following
condition:
\begin{equation}
\label{ineq: characterize thetahat 2}
	\nabla Q(\thb)^\top \thb \ = \ 0 \ \le \ \nabla Q(\thb)^\top \bs{e}
	\quad\text{for all} \ \bs{e} \in \EE \cup \{-\bs{1}\} .
\end{equation}
Thus we have to check only finitely many constraints. Note, however, that
the cardinality of $\EE$ may be substantially larger than the dimension
$p$, so that checking (\ref{ineq: characterize thetahat 2}) is far from
trivial.

\paragraph{Application to $\K_{r,s}$.}
Applying Lemma~\ref{lem: extremal set} to the cone $\K_{r,s} \subset
\R^{r\times s}$ yields the following representation: With
$$
	\EE_{r,s} \ = \ \K_{r,s} \cap \{0,1\}^{r\times s}
$$
any matrix $\x \in \K$ may be written as
$$
	\x \ = \ a_o \bs{1}_{r\times s}
		+ \sum_{\e \in \EE_{r,s}} \lambda_{\e} \e
$$
with coefficients $a_o \in \R$ and $\lambda_{\e} \ge 0$, $\e \in
\EE_{r,s}$.

There is a one-to-one correspondence between the set $\EE_{r,s}$ and the set of all vectors $\tilde{\bs{e}} \in \{1,2,\ldots,r+s\}^r$ with components $\tilde{e}_1 < \tilde{e}_2 < \cdots < \tilde{e}_r$ via the mapping
$$
	\bs{e} \ \mapsto \ \Bigl( i + \sum_{j=1}^s e_{ij} \Bigr)_{i=1}^r .
$$
Since such a vector $\tilde{\bs{e}}$ corresponds to a subset of $\{1,2,\ldots,r+s\}$ with $r$ elements, we end up with
$$
	\# \EE_{r,s} \ = \ \binom{r+s}{r} = \binom{r+s}{s} .
$$
Hence the cardinality of $\EE_{r,s}$ grows exponentially in $\min(r,s)$. Nevertheless, minimizing a linear functional over $\EE_{r,s}$ is possible in $O(rs)$ steps, as explained in the next section.

\paragraph{Proof of Lemma~\ref{lem: extremal set}.}
For $\x \in \K$ let $a_0 < a_1 < \cdots < a_m$ be the different elements of
$\{x_1, x_2, \ldots, x_p\}$, i.e.\ $a_0 = \min(\x)$ and $a_m = \max(\x)$.
Then
$$
	\x \ = \ a_0 \bs{1}
		+ \sum_{i=1}^m (a_i - a_{i-1}) \bigl( 1\{x_t \ge a_i\}
\bigr)_{t=1}^p .
$$
Obviously, these weights $a_i - a_{i-1}$ are nonnegative and sum to
$\max(\x) - \min(\x)$. Furthermore, one can easily deduce from $\x \in \K$ that $\bigl( 1\{x_t \ge a\} \bigr)_{t=1}^p$ belongs to $\EE$ for any real threshold $a$.	\hfill	$\Box$

\subsection{A dynamic program for $\bs{\EE_{r,s}}$}

For some matrix $\a \in \R^{r\times s}$ let $L : \R^{r\times s} \to \R$ be
given by
$$
	L(\x) \ = \ \sum_{i=1}^r\sum_{j=1}^s a_{ij} x_{ij} .
$$
The minimum of $L(\cdot)$ over the finite set $\EE_{r,s}$ may be obtained
by means of the following recursion: For $1 \le k \le r$ and $1 \le \ell
\le s$ define
\bea
	H(k,\ell) & = & \min \Bigl\{ \sum_{i=k}^r \sum_{j=1}^s
a_{ij}e_{ij} :
		\e \in \EE_{r,s}, e_{k \ell} = 1 \Bigr\} , \\
	H(k,s+1) & = & \min \Bigl\{ \sum_{i=k}^r \sum_{j=1}^s a_{ij}e_{ij}
:
		\e \in \EE_{r,s} \Bigr\} .
\eea
Then
$$
	\min_{\e \in \EE_{r,s}} L(\e) \ = \ H(1,s+1) ,
$$
and
\bea
	H(k,1)    & = & \sum_{i=k}^r \sum_{j=1}^s a_{ij} , \\
	H(k,\ell+1) & = & \min \Bigl( H(k,\ell),
			\sum_{j=\ell+1}^s a_{ij} + H(k+1,\ell+1) \Bigr)
\eea
where we use the conventions that $H(k+1,\cdot) = 0$ and $\sum_{j=s+1}^s
\cdot = 0$. In the recursion formula for $H(k,\ell+1)$, the term
$\sum_{j=\ell+1}^s a_{ij} + H(k+1,\ell+1)$ is the minimum of $L_k(\e) =
\sum_{i=k}^r \sum_{j=1}^s a_{ij} e_{ij}$ over all matrices $\e \in
\EE_{r,s}$ with $e_{k\ell} = 0$ and $e_{k,\ell+1} = 1$ (if $\ell < s$),
while $H(k,\ell)$ is the minimum of $L_k(\e)$ over all $\e \in \EE_{k,s}$
with $e_{k\ell} = 1$.

Table~\ref{tab: dynamic program} provides pseudocode for an algorithm that determines a minimizer of $L(\cdot)$ over $\EE_{r,s}$.

\begin{table}[h]
\centering
{\bf\begin{tabular}{|l|} \hline
\ruck{0}	Algorithm $\e \leftarrow \mbox{DynamicProgram}(\a)$\\
\ruck{0}	$\b \leftarrow \bigl( \sum_{j=\ell}^s a_{k,j} \bigr)_{k \le
r, \ell \le s+1}$\\
\ruck{0}	$\H \leftarrow \bigl(0\bigr)_{k \le r+1, \ell \le s+1}$\\
\ruck{0}	for $k \leftarrow r$ downto $1$ do\\
\ruck{1}		$H_{k,1} \leftarrow H_{k+1,1} + b_{k,1}$\\
\ruck{1}		for $\ell \leftarrow 1$ to $s$ do\\
\ruck{2}			$H_{k,\ell+1} \leftarrow
        				\min \bigl( H_{k,\ell},
				                b_{k,\ell+1} + H_{k+1,\ell+1} \bigr)$\\
\ruck{1}		end for\\
\ruck{0}	end for\\
\ruck{0}	$\e \leftarrow \bigl(0\bigr)_{k\le r, \ell\le s}$\\
\ruck{0}	$k \leftarrow 1$, $\ell \leftarrow s$\\
\ruck{0}	while $k \le r$ and $\ell \ge 1$ do\\
\ruck{1}		if $H_{k,\ell+1} = H_{k,\ell}$ then\\
\ruck{2}			$(e_{i,\ell})_{i=k}^r \leftarrow (1)_{i=k}^r$\\
\ruck{2}			$\ell \leftarrow \ell-1$\\
\ruck{1}		else\\
\ruck{2}			$k \leftarrow k+1$\\
\ruck{1}		end if\\
\ruck{0}	end while.\\\hline
\end{tabular}}
\caption{Minimizing a linear functional over $\EE_{r,s}$}
\label{tab: dynamic program}
\end{table}

\subsection{Active set type algorithms}

Throughout this exposition we assume that minimization of $Q$ over an
affine linear subspace of $\R^p$ is feasible. This is certainly the case if
$Q$ is a quadratic functional. If $Q$ is twice continuously differentiable
with positive definite Hessian matrix everywhere, this minimization problem
can be solved with arbitrarily high accuracy by a Newton type algorithm.

All algorithms described in this paper alternate between two basic
procedures which are described next. In both procedures $\thb \in \K$ is
replaced with a vector $\thb_{\rm new} \in \K$ such that $Q(\thb_{\rm new})
< Q(\thb)$ unless $\thb_{\rm new} = \thb$.

\subsubsection*{Basic procedure 1: Checking optimality of $\thb \in \K$}

Suppose that $\thb \in \K$ satisfies already the the following two
equations:
\begin{equation}
\label{eq: gradient condition}
	\nabla Q(\thb)^\top \thb \ = \ 0 \ = \ \nabla Q(\thb)^\top \bs{1} .
\end{equation}
According to (\ref{ineq: characterize thetahat}), this vector is already
the solution $\hat{\thb}$ if, and only if, $\nabla Q(\thb)^\top \bs{e} \ge
0$ for all $\bs{e} \in \EE$. Thus we determine
$$
	\Delta \ \in \ \argmin_{\bs{e} \in \EE} \, \nabla Q(\thb)^\top
\bs{e}
$$
and do the following: If $\nabla Q(\thb)^\top \Delta \ge 0$, we know that
$\thb = \hat{\thb}$ and stop the algorithm. Otherwise we determine
$$
	t_o \ = \ \argmin_{t \in \R} \, Q(\thb + t \Delta) \ > \ 0
$$
and replace $\thb$ with
$$
	\thb_{\rm new} \ := \ \thb + t_o \Delta .
$$
This vector $\thb_{\rm new}$ lies in the cone $\K$, too, and satisfies the
inequality $Q(\thb_{\rm new}) < Q(\thb)$. Then we proceed with basic
procedure 2.

\subsubsection*{Basic procedure 2:
Replacing $\thb \in \K$ with a ``locally optimal'' point $\thb_{\rm new}
\in \K$}

The general idea of basic procedure 2 is to find a point $\thb_{\rm new}
\in \K$ such that
\begin{equation}
\label{eq: local optimality}
	\thb_{\rm new} \ = \ \argmin_{\x \in \V} \, Q(\x)
\end{equation}
for some $\V$ in a finite family $\VV$ of linear subspaces of $\R^p$.
Typically these subspaces $\V$ are obtained by replacing some inequality
constraints from $\CC$ with equality constraints and ignoring the remaining
ones. This approach is described below as basic procedure 2a. But we shall
see that it is potentially useful to modify this strategy; see basic
procedures 2b and 2c.

\paragraph{Basic procedure 2a: The classical active set approach.}
For $\thb \in \K$ define
$$
	\V(\thb) \ = \ \bigl\{ \x \in \R^p : x_u = x_v \ \text{for all} \
(u,v) \in \CC
		\ \text{with} \ \th_u = \th_v \bigr\} .
$$
This is a linear subspace of $\R^p$ containing $\bs{1}$ and $\thb$ which is
determined by those constraints from $\CC$ which are ``active'' in $\thb$.
It has the additional property that for any vector $\x \in \V(\thb)$,
$$
	\lambda(\thb,\x)
		= \max \bigl\{ t \in [0,1] : (1 - t)\thb + t \x \in \K
\bigr\}
	\ > \ 0 .
$$
Precisely, $\lambda(\thb,\x) = 1$ if $\x \in \K$, and otherwise,
$$
	\lambda(\thb,\x) \ = \ \min_{(u,v) \in \CC \,:\, x_u > x_v} \,
		\frac{\th_v - \th_u}{\th_v - \th_u - x_v + x_u} .
$$

The key step in basic procedure 2a is to determine $\x_o = \argmin_{\x \in
\V(\thb)} Q(\x)$ and $\lambda(\thb,\x_o)$. If $\x_o \in \K$, which is
equivalent to $\lambda(\thb,\x_o) = 1$, we are done and return $\thb_{\rm
new} = \x_o$. This vector satisfies (\ref{eq: local optimality}) with $\V =
\V(\thb)$ and $\V = \V(\thb_{\rm new})$. The latter fact follows simply
from $\V(\thb_{\rm new}) \subset \V(\thb)$. If $\x_o \not\in \K$, we repeat
this key step with $\thb_{\rm new} = (1 - \lambda(\thb,\x_o)\thb +
\lambda(\thb,\x_o) \x_o$ in place of $\thb$.

In both cases the key step yields a vector $\thb_{\rm new}$ satisfying
$Q(\thb_{\rm new}) < Q(\thb)$, unless $\x_o = \thb$. Moreover, if $\x_o
\not\in \K$, then the vector space $\V(\thb_{\rm new})$ is contained in
$\V(\thb)$ with strictly smaller dimension, because at least one additional
constraint from $\CC$ becomes active. Hence after finitely many repetitions
of the key step, we end up with a vector $\thb_{\rm new}$ satisfying
(\ref{eq: local optimality}) with $\V = \V(\thb_{\rm new})$.
Table~\ref{tab: basic procedure 2a} provides pseudocode for basic procedure
2a.

\begin{table}[h]
\centering
{\bf\begin{tabular}{|l|}
\hline
\ruck{0}	Algorithm $\thb_{\rm new} \leftarrow
\mbox{BasicProcedure2a}(\thb)$\\
\ruck{0}	$\thb_{\rm new} \leftarrow \thb$\\
\ruck{0}	$\x_o \leftarrow \argmin_{\x \in \V(\thb_{\rm new})}
Q(\x)$\\
\ruck{0}	$\lambda \leftarrow \lambda(\thb_{\rm new}, \x_o)$\\
\ruck{0}	while $\lambda < 1$ do\\
\ruck{1}		$\thb_{\rm new} \leftarrow (1 - \lambda)\thb_{\rm
new} + \lambda\x_o$\\
\ruck{1}		$\x_o \leftarrow \argmin_{\x \in \V(\thb_{\rm
new})} Q(\x)$\\
\ruck{1}		$\lambda \leftarrow \lambda(\thb_{\rm new},
\x_o)$\\
\ruck{0}	end while\\
\ruck{0}	$\thb_{\rm new} \leftarrow \x_o$\\
\hline
\end{tabular}}
\caption{Basic procedure 2a}
\label{tab: basic procedure 2a}
\end{table}

\paragraph{Basic procedure 2b: Working with complete orders.}
The determination and handling of the subspace $\V(\thb)$ in basic
procedure 2a may be rather involved, in particular, when the set $\CC$
consists of more than $p$ constraints. One possibility to avoid this is to
replace $\V(\th)$ and $\K$ in the key step with the following subspace
$\V^*(\thb)$ and cone $\K^*(\thb)$, respectively:
\bea
	\V^*(\thb) & = & \bigl\{ \x \in \R^p : \text{for all} \ u,v \in
\{1,\ldots,p\}, \
		x_u = x_v \ \text{if} \ \th_u = \th_v \bigr\} , \\
	\K^*(\thb) & = & \bigl\{ \x \in \R^p : \text{for all} \ u,v \in
\{1,\ldots,p\}, \
		x_u \le x_v \ \text{if} \ \th_u \le \th_v \bigr\} .
\eea
Note that $\bs{1}, \thb \in \K^*(\thb) \subset \V^*(\thb)$, and one easily
verifies that $\K^*(\thb) \subset \K$ if $\thb \in \K$. Basic procedure 2b
works precisely like basic procedure 2a, but with $\V^*(\cdot)$ in place of
$\V(\cdot)$, and $\lambda(\thb,\x)$ is replaced with
$$
	\lambda^*(\thb,\x) \ = \ \max \bigl\{ t \in [0,1] :
		(1 - t)\thb + t\x \in \K^*(\thb) \bigr\} .
$$
Then basic procedure 2b yields a vector $\thb_{\rm new}$ satisfying
(\ref{eq: local optimality}) with $\V = \V^*(\thb_{\rm new})$.

When implementing this procedure, it is useful to determine a permutation
$\sig(\cdot)$ of $\{1,\ldots,p\}$ such that $\th_{\sig(1)} \le
\th_{\sig(2)} \le \cdots \le \th_{\sig(p)}$. Let $1 \le i_1 < i_2 < \cdots
< i_q = p$ denote those indices $i$ such that $\th_{\sig(i)} <
\th_{\sig(i+1)}$ if $i < p$. Then, with $i_0 = 0$,
\bea
	\V^*(\thb) & = & \bigl\{ \x \in \R^p :
		\text{for} \ 1 \le \ell \le q, \ \
		x_{\sig(i)} \ \text{is constant in} \ i \in
\{i_{\ell-1}+1,\ldots,i_\ell\}
		\bigr\}	, \\
	\K^*(\thb) & = & \bigl\{ \x \in \V^*(\thb) :
		\text{for} \ 1 \le \ell < q, \ \
		x_{\sig(i_{\ell})} \le x_{\sig(i_{\ell+1})} \bigr\} ,
\eea
and
$$
	\lambda^*(\thb,\x)
	\ = \ \min_{2 \le \ell \le p \,:\, x_{\sig(i_{\ell-1})} >
x_{\sig(i_\ell)}} \,
		\frac{ \th_{\sig(i_\ell)} - \th_{\sig(i_{\ell-1})} }
		     { \th_{\sig(i_\ell)} - \th_{\sig(i_{\ell-1})}
		      -  x_{\sig(i_\ell)} +   x_{\sig(i_{\ell-1})} } .
$$

\paragraph{Basic procedure 2c: A shortcut via the PAVA.}
In the special case of $Q(\thb)$ being the weighted least squares
functional in (\ref{eq: WLS}), one can determine
$$
	\thb_{\rm new} \ = \ \argmin_{\x \in \K^*(\thb)} \, Q(\x)
$$
directly by means of the PAVA with a suitable modification for the equality
constraints defining $\V^*(\thb)$.

\subsection*{The whole algorithm and its validity}

All subspaces $\V(\thb)$ and $\V^*(\thb)$, $\thb \in \K$, correspond to partitions of $\{1,2,\ldots,p\}$ into index sets. Namely, the linear subspace corresponding to such a partition consists of all vectors $\x \in \R^p$ with the property that $x_u = x_v$ for arbitrary indices $u,v$ belonging to the same set from the partition. Thus the subspaces used in basic procedures 2a-b belong to a finite family $\VV$ of linear subspaces of $\R^p$ all containing $\bs{1}$.

We may start the algorithm with initial point
$$
	\thb^{(0)} \ = \ \Bigl( \argmin_{t \in \R} Q(t \bs{1}) \Bigr)
\cdot \bs{1} .
$$
Now suppose that $\thb^{(0)}, \ldots, \thb^{(k)} \in \K$ have been chosen such that
$$
	\thb^{(\ell)} \ = \ \argmin_{\x \in \V^{(\ell)}} \, Q(\x)
	\quad\text{for} \ 1 \le \ell \le k
$$
with linear spaces $\V^{(0)}, \ldots, \V^{(k)} \in \VV$. Then $\thb =
\thb^{(k)}$ satisfies (\ref{eq: gradient condition}), and we may apply
basic procedure 1 to check whether $\thb^{(k)} = \hat{\thb}$. If not, we may also apply a variant of basic procedure 2 to get $\thb^{(k+1)} \in \K$ minimizing $Q$ on a linear subspace $\V^{(k+1)} \in \VV$, where $Q(\thb^{(k+1)}) < Q(\thb^{(k)})$. Since $\VV$ is finite, we will obtain $\hat{\thb}$ after finitely many steps.

Similar arguments show that our algorithm based on basic procedure~2c reaches an optimum after finitely many steps, too.

\paragraph{Final remark on coercivity.}
As mentioned for Setting~2a, the algorithm above may be applicable even in situations when the functional $Q$ fails to be coercive. In fact, we only need to assume that $Q$ attains a minimum, possibly non-unique, over any linear space $\V(\thb)$, $\V^*(\thb)$ or any cone $\K^*(\thb)$, and we have to able to compute it. In Setting~2a, one can verify this easily.

\section{Shrinkage estimation}
\label{sec: shrinkage}

We consider a regression setting as in Section~\ref{sec: WLS for bimonotone regression}, this time with Gaussian errors $\eps^t \sim \mathcal{N}(0, \sigma^2)$. As before, the regression function $\mu : \R \times \R \to \R$ is reduced to a matrix
$$
	\bs{M} = \bigl( \mu(x_{(i)},y_{(j)}) \bigr)_{i,\,j}^{}
	\in \R^{r\times s}
$$
for given design points $x_{(1)} < x_{(2)} < \cdots < x_{(r)}$ and $y_{(1)} < y_{(2)} < \cdots < y_{(s)}$. This matrix is no longer assumed to be bimonotone, but the latter constraint will play a role in our estimation method.

\subsection{Transforming the signal}

At first we represent the signal $\bs{M}$ with respect to a certain basis of $\R^{r\times s}$. To this end let $\bs{U} = [\bs{u}_1 \, \bs{u}_2 \, \ldots \, \bs{u}_r]$ and $\bs{V} = [\bs{v}_1 \, \bs{v}_2 \, \ldots \, \bs{v}_s]$ be orthonormal matrices in $\R^{r\times r}$ and $\R^{s\times s}$, respectively, to be specified later. Then we write
$$
	\bs{M}
	\ = \ \bs{U} \tilde{\bs{M}} \bs{V}^\top
	\ = \ \sum_{i,j} \tilde{M}_{ij}^{} \, \bs{u}_i^{} \bs{v}_j^\top
	\quad\text{with}\quad
	\tilde{\bs{M}}
	\ = \ \bs{U}^\top \bs{M} \bs{V}
	\ = \ \bigl( \bs{u}_i^\top \bs{M} \bs{v}_j^{} \bigr)_{i,j} .
$$
Thus $\tilde{\bs{M}}$ contains the coefficients of $\bs{M}$ with respect to the new basis matrices $\bs{u}_i^{}\bs{v}_j^\top \in \R^{r\times s}$. The purpose of such a transformation is to obtain a transformed signal $\tilde{\bs{M}}$ with many coefficients being equal or at least close to zero.

One particular construction of such basis matrices $\bs{U}$ and $\bs{V}$ is via discrete smoothing splines: For given degrees $k, \ell \ge 1$, consider \textsl{annihilators}
\bea
	\bs{A} & = & \begin{bmatrix}
		a_{11} & \cdots & a_{1,k+1} &             &        & 0 \\
		       & a_{22} & \cdots    & a_{2,k+2}   &        &   \\
		       &        & \ddots    &             & \ddots &   \\
		0      &        &           & a_{r-k,r-k} & \cdots & a_{r-k,r}
	\end{bmatrix} \ \in \ \R^{(r-k)\times r} , \\[1ex]
	\bs{B} & = & \begin{bmatrix}
		b_{11} & \cdots & b_{1,\ell+1} &                   &        & 0 \\
		       & b_{22} & \cdots       & b_{2,\ell+2}      &        &   \\
		       &        & \ddots       &                   & \ddots &   \\
		0      &        &              & b_{s-\ell,s-\ell} & \cdots & b_{s-\ell,s}
	\end{bmatrix} \ \in \ \R^{(s-\ell)\times s} ,
\eea
with unit row vectors such that
\bea
	\bs{A} \bigl( x_{(i)}^e \bigr)_{i=1}^r & = & \bs{0}
		\quad\text{for} \ e=0,\ldots,k-1 , \\
	\bs{B} \bigl( y_{(j)}^e \bigr)_{j=1}^s & = & \bs{0}
		\quad\text{for} \ e=0,\ldots,\ell-1 .
\eea
An important special case is $k = \ell = 1$. Here
$$
	\bs{A} \ = \ \frac{1}{\sqrt{2}} \begin{bmatrix}
		1 & -1 &        &        &  0 \\
		  &  1 & -1     &        &    \\
		  &    & \ddots & \ddots &    \\
		0 &    &        & 1      & -1
	\end{bmatrix}
	\quad\text{and}\quad
	\bs{B} \ = \ \frac{1}{\sqrt{2}} \begin{bmatrix}
		1 & -1 &        &        &  0 \\
		  &  1 & -1     &        &    \\
		  &    & \ddots & \ddots &    \\
		0 &    &        & 1      & -1
	\end{bmatrix}
$$
satisfy the equations $\bs{A} \bs{1}_r = \bs{0}$ and $\bs{B} \bs{1}_s =
\bs{0}$.

Next we determine singular value decompositions of $\A$ and $\B$, namely,
\bea
	\A & = & \tilde{\bs{U}} \cdot
		\bigl[ \bs{0}_{(r-k) \times k}^{} \,
		       \underbrace{\mathrm{diag}(a_1, \ldots, a_{r-k})
		                   }_{0 \ \le \ a_1 \ \le \ \cdots \ \le \ a_{r-k}} \bigr]
			\cdot \bs{U}^\top \\
	\B & = & \tilde{\bs{V}} \cdot
		\bigl[ \bs{0}_{(s-\ell) \times \ell}^{} \,
		       \underbrace{\mathrm{diag}(b_1, \ldots, b_{s-\ell})
		                   }_{0 \ \le \ b_1 \ \le \ \cdots \ \le \ b_{s-\ell}} \bigr]
			\cdot \bs{V}^\top
\eea
with column-orthonormal matrices $\tilde{\bs{U}}$, $\bs{U} = [\bs{u}_1 \, \bs{u}_2 \, \cdots \, \bs{u}_r]$, $\tilde{\bs{V}}$ and $\bs{V} = [\bs{v}_1 \, \bs{v}_2 \, \cdots \, \bs{v}_s]$. The vectors $\bs{u}_1,\ldots,\bs{u}_k$ and $\bs{v}_1,\ldots,\bs{v}_\ell$ correspond to the space of polynomials of
order at most $k$ and $\ell$, respectively. In particular, we always choose $\bs{u}_1 = r^{-1/2} \bs{1}_r$ and $\bs{v}_1 = s^{-1/2} \bs{1}_s$. Then
$$
	\begin{array}{rccl}
	\bs{M}
	& = & \tilde{M}_{11}^{} \, \bs{u}_1^{} \bs{v}_1^\top
			& \quad (\text{constant part}) \\
	&& \displaystyle
		+ \ \sum_{i=2}^r \tilde{M}_{i1}^{} \,
			\bs{u}_i^{}\bs{v}_1^\top
		+ \sum_{j=2}^s \tilde{M}_{1j}^{} \,
			\bs{u}_1^{}\bs{v}_j^\top
			& \quad (\text{additive part}) \\
	&& \displaystyle
		+ \ \sum_{i,j \ge 2} \tilde{M}_{ij}^{} \,
			\bs{u}_i^{}\bs{v}_j^\top
			& \quad (\text{interactions})
	\end{array}
$$
One may also write
$$
	\bs{M}
	\ = \ \bs{U} \ \begin{array}{|c|c|}
		\hline
		\text{polynomial part} & \text{half-polyn.\ interactions}
		\\
		k \times \ell          & k \times (s - \ell) \\
		\hline
		\text{half-polyn.\ interactions} & \text{non-polyn.\ interactions}
		\\
		(r - k) \times \ell & (r - k) \times (s - \ell) \\
		\hline
	\end{array} \ \bs{V}^\top .
$$
For moderately smooth functions $\mu$ we expect $|\tilde{M}_{ij}|$ to have a decreasing trend in $i > k$ and in $j > \ell$. This motivates a class of shrinkage estimators which we describe next.

\subsection{Shrinkage estimation in the simple balanced case}

In the case of $n = p = rs$ observations such that each grid point $(x_{(i)},y_{(j)})$ is contained in $\bigl\{ (x^1,y^1), \ldots, (x^n,y^n) \bigr\}$, our input data may be written as a matrix
$$
	\bs{Z} \ = \ \bs{M} + \bs{\eps}
$$
with $\bs{\eps} \in \R^{r \times s}$ having independent components $\eps_{ij} \sim \mathcal{N}(0, \sigma^2)$. Reexpressing such data with respect to the discrete spline basis leads to $\tilde{\bs{Z}} = \tilde{\bs{M}} + \tilde{\bs{\eps}}$ with $\tilde{\bs{Z}} := \bs{U}^\top \bs{Z} \bs{V}$ and $\tilde{\bs{\eps}} := \bs{U}^\top \bs{\eps} \bs{V}$. Note that the raw data $\bs{Z}$ is the maximum likelihood estimator of $\bs{M}$. To benefit from the bias-variance trade-off, we consider component-wise shrinkage of the coefficient matrix $\tilde{\bs{Z}}$: For $\bs{\gamma} \in [0,1]^{r\times s}$ we consider the candidate estimator
\begin{equation}
	\label{eq: candidate}
	\hat{\bs{M}}^{(\bs{\gamma})}
	\ = \ \bs{U} \, (\gamma_{ij} \tilde{Z}_{ij})_{i,j}^{} \, \bs{V}^\top .
\end{equation}
Eventually we will choose a shrinkage matrix $\hat{\bs{\gamma}}$ depending on the data and compute the shrinkage estimator
\begin{equation}
	\label{eq: shrinkage estimator}
	\hat{\bs{M}} \ = \ \hat{\bs{M}}^{(\hat{\bs{\gamma}})} .
\end{equation}

Let $\|\bs{A}\|_F$ denote the Frobenius norm of a matrix $\bs{A}$, i.e.\ $\|\bs{A}\|_F^2 = \sum_{i,j} A_{ij}^2 = \mathrm{trace}(\bs{A}^\top\bs{A})$. As a measure of risk of the estimator (\ref{eq: candidate}), we consider
\bea
	R(\bs{\gamma}, \bs{M})
	& = & \Ex \bigl\| \hat{\bs{M}}^{(\gamma)} - \bs{M} \bigr\|_F^2 \\
	& = & \sum_{i,j} \bigl(  (1 - \gamma_{ij})^2 \tilde{M}_{ij}^2
		+ \sigma^2 \gamma_{ij}^2 \bigr) \\
	& = & \sum_{i,j} (\tilde{M}_{ij}^2 + \sigma^2)
		\Bigl( \gamma_{ij} - \frac{\tilde{M}_{ij}^2}{\tilde{M}_{ij}^2 + \sig^2} \Bigr)^2
			+ \sum_{i,j} \frac{\tilde{M}_{ij}^2 \sigma^2}{\tilde{M}_{ij}^2 + \sigma^2} .
\eea
Here we used the fact that the transformed error matrix $\tilde{\bs{\eps}}$ has the same distribution as $\bs{\eps}$. An estimator of this risk is given by
\bea
	\hat{R}(\bs{\gamma})
	& = & \sum_{i,j} \bigl( \hat{\sigma}^2 \gamma_{ij}^2
		+ (1 - \gamma_{ij})^2 (\tilde{Z}_{ij}^2 - \hat{\sigma}^2) \bigr) \\
	& = & \sum_{i,j} \tilde{Z}_{ij}^2
			\bigl( \gamma_{ij} - (1 - \hat{\sigma}^2/\tilde{Z}_{ij}^2) \bigr)^2
		+ \sum_{i,j} \hat{\sigma}^2
			\bigl( 1 - \hat{\sigma}^2/\tilde{Z}_{ij}^2 \bigr) ,
\eea
where $\hat{\sigma}$ is a certain estimator of $\sigma$, e.g.\ based on
high frequency components of $\tilde{\bs{Z}}$, see later.

Thus optimal shrinkage factors would be given by $\check{\gamma}_{ij} = \tilde{M}_{ij}^2/(\tilde{M}_{ij}^2 + \sigma^2)$, but these depend on the unknown signal $\bs{M}$. Naive estimators would be $\hat{\gamma}_{ij} = (1 - \hat{\sigma}^2/\tilde{Z}_{ij}^2)^+$. The resulting estimator's performance is rather poor, but it improves substantially if $\hat{\bs{\gamma}}$ in (\ref{eq: shrinkage estimator}) is given by
\begin{equation}
\label{eq: thresholding}
	\hat{\gamma}_{ij} \ = \ \max \Bigl( 1
		- \frac{\tau \log(p) \hat{\sigma}^2}
		       {\tilde{Z}_{ij}^2}, \, 0 \Bigr)
\end{equation}
with $\tau$ close to $2$; cf.\ Donoho and Johnstone \cite{donoho_johnstone_1994}.

An alternative strategy, utilized for instance by Beran and D\"umbgen \cite{beran_duembgen_1998}, is to restrict $\bs{\gamma}$ to a certain convex set of shrinkage matrices serving as a caricature of the optimal $\bs{\gamma}$. The previous considerations suggest to restrict $-\bs{\gamma}$ to be contained in $\K_{r,s}^{(k,\ell)}$, the set of all matrices $\thb \in \R^{r\times s}$ such that

\noindent
$\bullet$ \ $\th_{1,j} = \th_{2,j} = \cdots = \th_{k,j}$ is non-decreasing in $j > \ell$,\\
$\bullet$ \ $\th_{i,1} = \th_{i,2} = \cdots = \th_{i,\ell}$ is non-decreasing in $i > k$,\\
$\bullet$ \ $(\th_{ij})_{i > k, j > \ell}$ belongs to $\mathbb{K}_{r-k, s - \ell}$.

The set of all such shrinkage matrices $\bs{\gamma}$ is denoted by $\mathbb{G}_{r,s}^{(k,\ell)} = (- \K_{r,s}^{(k,\ell)}) \cap [0,1]^{r\times s}$. Thus we propose to use the shrinkage matrix
\begin{equation}
\label{eq: bimonotone shrinkage}
	\hat{\bs{\gamma}}
	\ = \ \argmin_{\bs{\gamma} \in \mathbb{G}_{r,s}^{(k,\ell)}} \,
		\hat{R}(\bs{\gamma}) .
\end{equation}
In the present setting one can show (cf.\ \cite{beran_duembgen_1998}) that
\bea
	\check{\bs{\gamma}}
		\ = \ \argmin_{\bs{\gamma} \in \mathbb{G}_{r,s}^{(k,\ell)}} \,
			R(\bs{\gamma}, \bs{M})
	& = & \Bigl( \frac{\check{\eta}_{ij}}{\check{\eta}_{ij} + \sig^2} \Bigr)_{i,j} \\
	\text{with}\quad
	\check{\etab}
	& = & - \, \argmin_{\thb \in \K_{r,s}^{(k,\ell)}}
		\sum_{i,j} \bigl( - (\tilde{M}_{ij}^2 + \sigma^2) - \th_{ij} \bigr)^2 .
\eea
Similarly,
\bea
	\hat{\bs{\gamma}}
		\ = \ \argmin_{\bs{\gamma} \in \mathbb{G}_{r,s}^{(k,\ell)}} \,
			\hat{R}(\bs{\gamma})
	& = & \bigl( (1 - \hat{\sig}^2/\hat{\eta}_{ij})^+ \bigr)_{i,j} \\
	\text{with}\quad
	\hat{\etab}
	& = & - \, \argmin_{\thb \in \K_{r,s}^{(k,\ell)}}
		\sum_{i,j} (- \tilde{Z}_{ij}^2 - \th_{ij})^2 .
\eea
This allows one to experiment with different values for $\hat{\sigma}$ with
little effort.

\paragraph{Estimation of the noise level.}
Two particular estimators are given by
\begin{equation}
\label{eq: shat_o}
	\hat{\sigma}_{1,\kappa}^{} \ = \ \biggl(
		\frac{\sum_{i/r + j/s \ge \kappa} \tilde{Z}_{ij}^2}
		     {\# \{(i,j) : i/r + j/s \ge \kappa\}} \biggr)^{1/2}
	\ \text{or}\quad
	\hat{\sigma}_{2,\kappa}^{}
	\ = \ \frac{\text{Median} \bigl( |\tilde{Z}_{ij}| : i/r + j/s \ge \kappa \bigr)}
	           {\Phi^{-1}(3/4)}
\end{equation}
for a certain number $\kappa \in (0,2)$, where $\Phi^{-1}$ denotes the standard Gaussian quantile function. The idea is that for $i >> 1$ and $j >> 1$, the components $\tilde{Z}_{ij}$ are essentially equal to the noise variables $\tilde{\eps}_{ij} \sim \mathcal{N}(0,\sigma^2)$. Otherwise both estimators tend to overestimate $\sigma$.

As to the choice of $\kappa$, we propose to choose it via visual inspection of the graphs of $\kappa \mapsto \hat{\sigma}_{1,\kappa}$ and $\kappa \mapsto \hat{\sigma}_{2,\kappa}$. Typically these functions are almost constant and close to $\sigma$ on a large subinterval of $(0,2)$, non-increasing to the left of that interval, and show random fluctuations to the right. As we shall illustrate later, the quality of the shrinkage estimator is rather robust with respect to the estimator $\hat{\sigma}$. In particular, overestimating $\sigma$ slightly is typically harmless or even beneficial.

\paragraph{Consistency.} 
We now augment the foregoing discussion with consistency results that follow from more general considerations in \cite{beran_duembgen_1998}. First of all, for large $p$, the normalized quadratic loss $p^{-1} \|\hat{\bs{M}}^{(\bs{\gamma})} - \bs{M}\|_F^2$ of a candidate estimator is close to its normalized risk $p^{-1} R(\bs{\gamma},\bs{M})$, uniformly over $\bs{\gamma} \in \mathbb{G}_{r,s}^{(k,\ell)}$. Precisely,
$$
	\Ex \sup_{\bs{\gamma} \in \mathbb{G}_{r,s}^{(k,\ell)}}
		\bigl| p^{-1} \|\hat{\bs{M}}^{(\bs{\gamma})} - \bs{M}\|_F^2
			- p^{-1} R(\bs{\gamma},\bs{M}) \bigr|
	\ \le \ C \, \frac{\sigma^2 + \sigma p^{-1/2} \|\bs{M}\|_F}{\max(r,s)^{1/2}}
$$
with $C$ denoting a generic universal constant. Moreover, if the variance estimator $\hat{\sigma}^2$ is $L_1$--consistent, the normalized estimated risk $p^{-1}\hat{R}(\bs{\gamma})$ differs little from the normalized true risk $p^{-1} R(\bs{\gamma},\bs{M})$, uniformly in $\bs{\gamma} \in \mathbb{G}_{r,s}^{(k,\ell)}$. Namely,
$$
	\Ex \sup_{\bs{\gamma} \in \mathbb{G}_{r,s}^{(k,\ell)}}
		\bigl| p^{-1} \hat{R}(\bs{\gamma}) - p^{-1} R(\bs{\gamma},\bs{M}) \bigr|
	\ \le \ C \, \frac{\sigma^2 + \sigma p^{-1/2} \|\bs{M}\|_F}{\max(r,s)^{1/2}}
		+ C \, \Ex \bigl| \hat{\sigma}^2 - \sigma^2 \bigr| .
$$
In particular, the shrinkage matrix $\hat{\bs{\gamma}}$ in (\ref{eq: bimonotone shrinkage}) and the corresponding estimator $\hat{\bs{M}} = \hat{\bs{M}}^{(\hat{\bs{\gamma}})}$ satisfy the inequalities
$$
	\left.\begin{array}{r}
		\Ex \bigl| p^{-1} \hat{R}(\hat{\bs{\gamma}}) - p^{-1} R_{\rm min}(\bs{M}) \bigr| \\[1ex]
		\Ex \bigl| p^{-1} \|\hat{\bs{M}} - \bs{M}\|_F^2 - p^{-1} R_{\rm min}(\bs{M}) \bigr|
	\end{array}\right\}
	\ \le \ C \, \frac{\sigma^2 + \sigma p^{-1/2} \|\bs{M}\|_F}{\max(r,s)^{1/2}}
		+ C \, \Ex \bigl| \hat{\sigma}^2 - \sigma^2 \bigr| ,
$$
where $R_{\rm min}(\bs{M})$ denotes the minimum of $R(\bs{\gamma},\bs{M})$ over all $\bs{\gamma} \in \mathbb{G}_{r,s}^{(k,\ell)}$.

\begin{Example}
\label{ex: Splash}
We generated a random matrix $\bs{Z} \in \R^{r\times s}$ with $r = 60$
rows, $s = 100$ columns and independent components $Z_{ij} \sim \mathcal{N}
\bigl( \mu(x_{(i)},y_{(j)}), 1 \bigr)$, where $x_{(i)} = (i - 0.5)/r$,
$y_{(j)} = (j - 0.5)/s$, and
$$
	\mu(x,y) \ = \ 2 \tau(x,y)^{-0.25} \sin(\tau(x,y)) + 0.05(x + y) ,
\quad
	\tau(x,y) \ = \ \sqrt{3x^2 + 2xy + 3y^2} + 1 .
$$
We smoothed this data matrix $\bs{Z}$ as described above with annihilators of order $k = \ell = 2$. The estimators $\hat{\sigma}_{1,\kappa}$ and $\hat{\sigma}_{2,\kappa}$ turned out to be almost constant and slightly smaller than $1.0$ on $(0.5, 0.65)$, so we chose $\hat{\sigma} = 1$. The first row of Figure~\ref{fig: splash 1} shows gray scale images of the raw data $\bs{Z}$ (left) and the true signal $\bs{M}$ (right). The second and third row depict the matrix $\hat{\bs{M}}$ for different values of $\hat{\sigma}$. Precisely, to show the effect of varying the estimated noise level, we replaced $\hat{\sigma}$ with $c \hat{\sigma}$, where $c = 0.5$ (undersmoothing), $c = 1.0$ (original estimator), $c = 1.5$ (oversmoothing) and $c = 2.0$ (heavy oversmoothing). In these pictures the gray scale ranges from $-7$ (black) to $7$ (white).

\begin{figure}[h]
\includegraphics[width=0.49\textwidth]{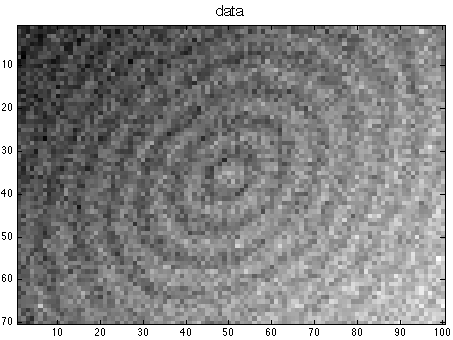}
\includegraphics[width=0.49\textwidth]{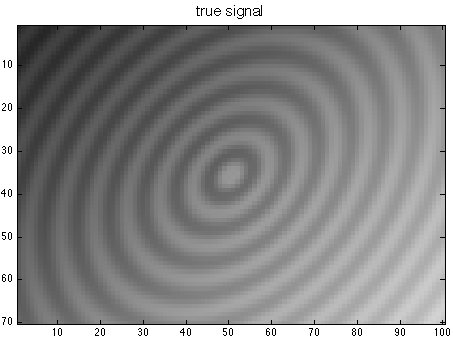}\\
\includegraphics[width=0.49\textwidth]{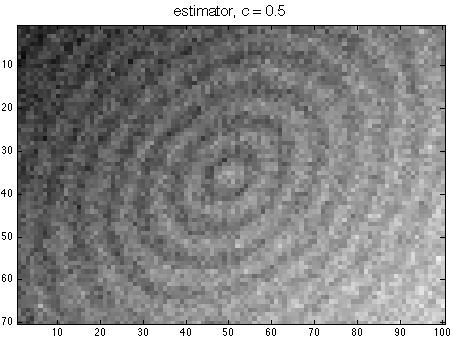}
\includegraphics[width=0.49\textwidth]{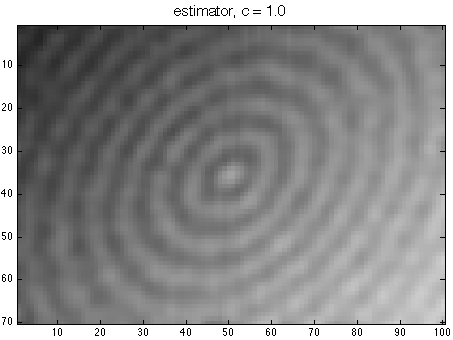}\\
\includegraphics[width=0.49\textwidth]{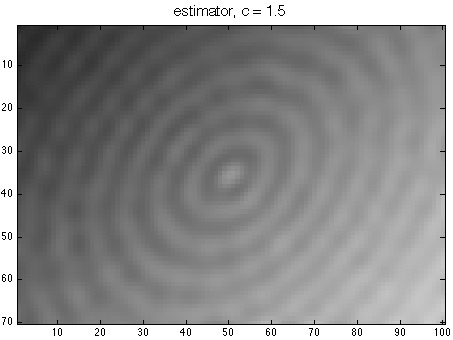}
\includegraphics[width=0.49\textwidth]{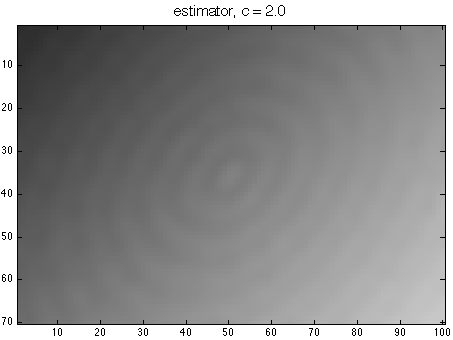}
\caption{Shrinkage estimation: data and true signal (1st row), estimators with $\hat{\sigma} \leftarrow c \hat{\sigma}$ for $c = 0.5, 1.0, 1.5, 2.0$ (2nd and 3rd row).}
\label{fig: splash 1}
\end{figure}

Figure~\ref{fig: splash 2} depicts the transformed squared coefficients $\tilde{Z}_{ij}^2 / (1 + \tilde{Z}_{ij}^2)$ (left panel) and the bimonotone shrinkage matrix $\hat{\bs{\gamma}}$ (right panel).

\begin{figure}[h]
\includegraphics[width=0.49\textwidth]{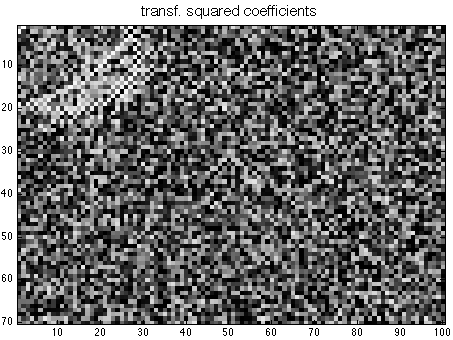}
\includegraphics[width=0.49\textwidth]{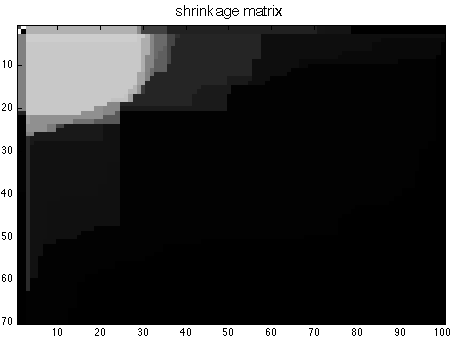}
\caption{Shrinkage estimation: Transformed squared coefficients $\tilde{Z}_{ij}^2 / (1 + \tilde{Z}_{ij}^2)$ (left) and bimonotone shrinkage matrix $\hat{\bs{\gamma}}$ (right).}
\label{fig: splash 2}
\end{figure}

Figure~\ref{fig: splash 3} shows the average squared loss $p^{-1} \|\hat{\bs{M}} - \bs{M}\|_F^2$ as a function of $\hat{\sigma}$. The emerging pattern is very stable over all simulations we looked at. This plot and figure~\ref{fig: splash 2} show that there is a rather large range of values for $\hat{\sigma}$ leading to estimators of similar quality. Overestimation of $\hat{\sigma}$ is less severe than underestimation and sometimes even beneficial.

\begin{figure}[h]
\centering
\includegraphics[width=0.6\textwidth]{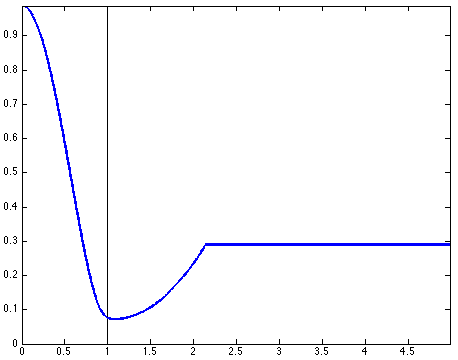}
\caption{Shrinkage estimation: Average quadratic loss as a function of
$\hat{\sigma}$.}
\label{fig: splash 3}
\end{figure}

Since this is just one simulation, we also conducted a simulation study. We generated 5000 such data matrices $\bs{Z}$. Each time we estimated the noise level via $\hat{\sigma} = \hat{\sigma}_{1,1}$. Then we computed the shrinkage estimators $\hat{\bs{M}}$ in (\ref{eq: shrinkage estimator}), where the shrinkage matrices $\hat{\bs{\gamma}}$ were given by (\ref{eq: bimonotone shrinkage}) and by (\ref{eq: thresholding}) with $\tau$ running through a fine grid of points in $(0,2]$. It turned out that $\tau = 0.60$ yielded optimal performance, although this value depends certainly on the underlying signal and noise level. Table~\ref{tab: MC risks} provides Monte Carlo estimates of the corresponding risk, i.e.\ the expectation of the normalized quadratic loss $p^{-1} \|\bs{\hat{\bs{M}}} - \bs{M}\|_F^2$. The values in brackets are the estimated standard deviations of the latter loss. This table shows that bimonotone shrinkage yields better results than componentwise (soft) thresholding.

\begin{table}[h]
$$
	\begin{array}{|c||c|c|c|c|c|}
	\hline
		\multicolumn{1}{|l||}{\text{bimonotone}}
		& \multicolumn{5}{l|}{\text{componentwise thresholding (\ref{eq: thresholding}) with}} \\
		\multicolumn{1}{|l||}{\text{shrinkage (\ref{eq: bimonotone shrinkage})}}
		& \tau = 0.5 & \tau = 0.6 & \tau = 1.0 & \tau = 1.5 & \tau = 2.0 \\
	\hline
		 0.0790  &  0.0922  &  0.0888  &  0.1044  &  0.1342  &  0.1619  \\
		(0.0044) & (0.0050) & (0.0051) & (0.0061) & (0.0073) & (0.0082) \\
	\hline
	\end{array}
$$
\caption{Estimated risks of different estimators in Example~\ref{ex: Splash}.}
\label{tab: MC risks}
\end{table}
\end{Example}

\subsection{Viticultural case study} 

\begin{figure}[h]
\includegraphics[width=0.49\textwidth]{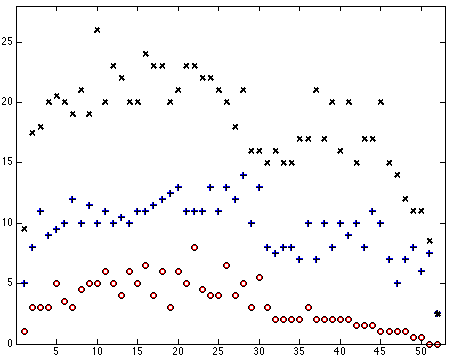}
\hfill
\includegraphics[width=0.49\textwidth]{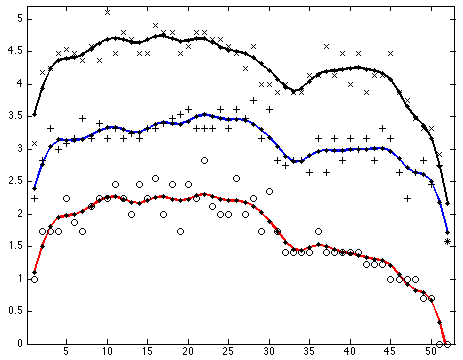}
\includegraphics[width=0.49\textwidth]{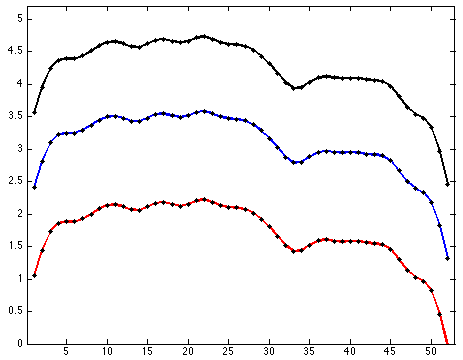}
\hfill
\includegraphics[width=0.49\textwidth]{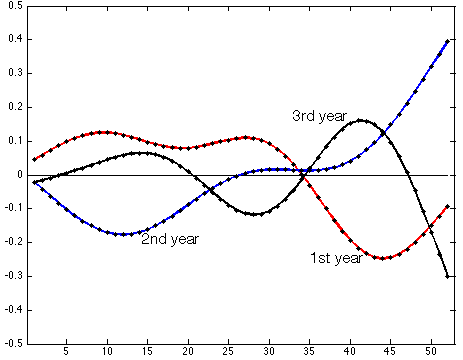}
\caption{Raw vineyard data (top left), transformed data and fitted values (top right), additive part (bottom left) and interactions (bottom right).}
\label{fig: Vineyard}
\end{figure}

In this case study, row $i$ of the data matrix $\bs{Y} \in \R^{52 \times 3}$ reports the grape yields harvested in $3$ successive years from a vineyard near Lake Erie that has $52$ rows of vines. The data is taken from Chatterjee, Handcock, and Simonoff \cite{chatterjee_etal_1995}. The grape yields, measured in lugs of grapes harvested from each vineyard-row, are plotted in the upper left panel of Figure~\ref{fig: Vineyard}, using a different plotting character for each of the three years. The analysis seeks to bring out patterns in the vineyard-row yields that persist across years. Year and vineyard-row are both ordinal covariates. The covariate vineyard-row summarizes location-dependent effects that may be due to soil fertility and microclimate. The covariate year summarizes time-varying effects that may be due to rainfall pattern, temperatures, and viticultural practices.

A preliminary data analysis based on running means and variance estimates from triplets $(Y_{i,j},Y_{i+1,j}, Y_{i+2,j})$, $1 \le i \le 50$, revealed that a square-root transformation yields a data matrix $\bs{Z} \in \R^{52 \times 3}$ which may be viewed as a two-way layout in which both the row and column numbers are ordinal covariates, the measurement errors are independent with mean zero and common unknown variance $\sigma^2$ and unknown mean matrix $\bs{M} = \Ex \bs{Z}$.

Now we applied the orthonormal transformation into spline bases with $x_{(i)} = i$ and $y_{(j)} = j$, where $k = 2$ and $\ell = 1$. In particular, $\bs{u}_1$ and $\bs{u}_2$ are proportional to $\bs{1}_{52}$ and $(i - 26.5)_{i=1}^{52}$, respectively. Similarly, $\bs{v}_1$, $\bs{v}_2$ and $\bs{v}_3$ are proportional to $\bs{1}_3$, $(-1, 0, 1)^\top$ and $(1, -2, 1)^\top$, respectively. The graphs of $\kappa \mapsto \hat{\sigma}_{1,\kappa}$ and $\kappa \mapsto \hat{\sigma}_{2,\kappa}$ revealed that $\hat{\sigma} = 0.25$ is a plausible estimator for $\sigma$. The resulting fitted matrix $\hat{\bs{M}}$ is shown in the upper right panel of Figure~\ref{fig: Vineyard}, adding linear interpolation between adjacent elements to bring out their trend. In addition the transformed data $Z_{ij}$ are superimposed as single points.

The estimated mean grape yields reveal shared patterns across the three years. Large dips in estimated mean grape yields occur in the outermost rows of the vineyard and near row $33$. These point to possible geographical variations in growing conditions, such as harsher climate at the vineyard edges or changes in soil fertility.  

It is also interesting to split the fit $\hat{\bs{M}}$ into an additive part (including constant) and interactions,
\bea
	\hat{\bs{M}}_{\rm add}^{}
	& = & \hat{\gamma}_{11}^{} \tilde{Z}_{11}^{} \, \bs{u}_1^{} \bs{v}_1^\top
		+ \sum_{i=2}^r \hat{\gamma}_{i1}^{} \tilde{Z}_{i1}^{} \, \bs{u}_i^{} \bs{v}_1^\top
		+ \sum_{j=2}^s \hat{\gamma}_{1j}^{} \tilde{Z}_{1j}^{} \, \bs{u}_1^{} \bs{v}_j^\top , \\
	\hat{\bs{M}}_{\rm inter}^{}
	& = & \sum_{i=2}^r \sum_{j=2}^s
		\hat{\gamma}_{ij}^{} \tilde{Z}_{ij}^{} \, \bs{u}_i^{} \bs{v}_j^\top .
\eea
The lower panels of Figure~\ref{fig: Vineyard} depict these parts separately. The plot of the additive part emphasizes the pattern across rows just described and the (nonlinear) increase across years. The interactions reveal that a simple additive model doesn't seem appropriate for these data.

\paragraph{Acknowledgements.}
This work was supported by the Swiss National Science Foundation. We are grateful to two reviewers for constructive comments.




\end{document}